\documentclass{gmee2012}
\usepackage{amsmath}
\usepackage{cite}     
\usepackage{graphicx}
\usepackage[load=accepted]{siunitx}
\usepackage{color}
\usepackage{psfrag}
\usepackage{pstricks}
\usepackage{subfigure}
\usepackage{booktabs}
\usepackage{url}
\usepackage{multirow}

\DeclareMathOperator{\varRe}{Re}
\DeclareMathOperator{\varIm}{Im}

\begin{document}
%
\title{\LARGE{A three-arm current comparator bridge, \\ for impedance comparisons over the complex plane}}



%
\author{Luca Callegaro, Vincenzo D'Elia, Massimo Ortolano, and Faranak Pourdanesh
\thanks{Luca Callegaro, Vincenzo D'Elia and Faranak Pourdanesh are with the Electromagnetism Division of the Istituto Nazionale di Ricerca Metrologica (INRIM), strada delle Cacce 91, 10135 Torino, Italy. E-mail: \texttt{l.callegaro@inrim.it}.} 
\thanks{Massimo Ortolano is with the Politecnico di Torino, Torino, Italy.}
}


\maketitle

\begin{abstract}
We present here the concept of three-arm current comparator impedance bridge, which allows comparisons among three unlike impedances. Its purpose is the calibration of impedances having arbitrary phase angles, against calibrated nearly-pure impedances. An analysis of the bridge optimal setting and proper operation is presented. To test the concept, a two terminal-pair digitally-assisted bridge has been realized; measurements of an air-core inductor and of an $RC$ network versus decade resistance and capacitance standards, at \si{\kilo\hertz} frequency, have been performed. The bridge measurements are compatible with previous knowledge of the standards' values with relative deviations in the \numrange{E-5}{E-6} range.
\end{abstract}

%
%
\section{Introduction}
Transformer ratio bridges are the workhorses of primary impedance metrology~\cite{Awan2010,CallegaroBook}; in this kind of bridges an impedance ratio is compared with a voltage or a current ratio standard~\cite{Kusters1964,Kusters1965,Moore1988}. The most accurate current ratio bridges are based on the \emph{current comparator} (CC) principle~\cite{Moore1988}. Typically, CC bridges have two main arms and are employed for comparing like\footnote{That is, having similar phase angles, such as two resistors or two capacitors.} impedances. 

We introduce here the concept of \emph{three-arm current comparator bridge}~\cite{Callegaro2014_CPEM3arm}, where three unlike impedances are involved in the measurement. The measurement outcome gives a relation between the complex values of the three impedances. The aim of the three-arm comparator is the calibration of impedances having arbitrary phase angles, with traceability to pure\footnote{That is, having phase angles near \si{0}{\degree} or $\pm \SI{90}{\degree}$.} impedances, as are the resistance and capacitance scales maintained in national metrology institutes and calibration centers.
\section{Three-arm current comparator bridge}
\subsection{Principle}
\begin{figure}[tb]
	\centering
	\def\svgwidth{5.9cm}	
	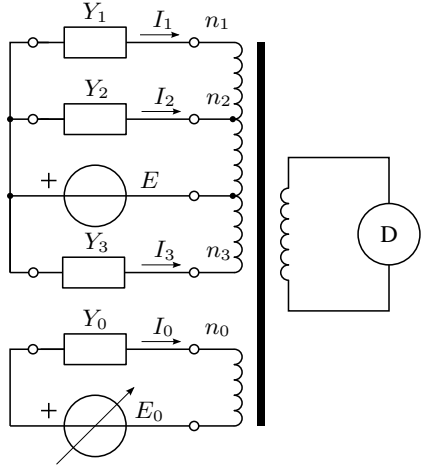
 	\caption{Three-arm current comparator. The voltage source $E$ feeds the admittances $Y_k$, $k=1,2,3$, which deliver currents $I_k$ to taps $n_k$ of the main winding. In the diagram, $n_1, n_2 >0$; $n_3 <0$. The injection current $I_0$, generated by the voltage $E_0$ and the admittance $Y_0$, flows through the injection winding having $n_0$ turns. The bridge equilibrium is sensed by the detector D connected to the detection winding. \label{fig:3armCC}}%
\end{figure}

The schematic diagram illustrating the principle of operation of the three-arm CC is shown in Fig.~\ref{fig:3armCC}. The bridge comprises three main arms, numbered $1,2,3$, and an injection arm $0$, which is employed to balance the bridge. Bridge equilibrium is sensed by the detector D. 

Each arm $k=0,\ldots,3$ includes an admittance $Y_k$ (or impedance $Z_k=Y_k^{-1}$) which, when excited, is crossed by current $I_k$. Current $I_k$ flows in the $n_k$ tap (chosen from an available tap set $\mathsf{T}$) of the primary windings of the current comparator. Arms $1,2,3$ are excited with the same fixed voltage $E$, whereas the injection arm $0$ is excited with an adjustable voltage $E_0$.

The bridge equilibrium is achieved when the sum of all magnetomotive forces $\mathcal{M}_k$ generated by the currents $I_k$ in the comparator core is nulled, that is, when 
\begin{equation}
	\label{eq:Mnull}
	0 = \sum_{k=0}^{3} \mathcal{M}_k = \sum_{k=0}^{3} n_k I_k = n_0 Y_0 E_0 + \left( \sum_{k=1}^{3} n_k Y_k \right) E\,.
\end{equation}
Assuming that $Y_1$ and $Y_2$ are calibrated standards, and that $Y_3$ is the admittance under measurement, \eqref{eq:Mnull} can be rewritten as the measurement model
\begin{equation}
	\label{eq:equilibrium}
	Y_3 = -\frac{1}{n_3} \left(n_1 Y_1 + n_2 Y_2 + n_0 Y_0 \frac{E_0}{E}\right).
\end{equation}

The bridge accuracy is mainly determined by the ratio accuracy of its current comparator. The injection $\mathcal{M}_0$ should be small with respect to the main magnetomotive forces $\mathcal{M}_k, \, k=1,2,3$, and the best accuracy is achieved for null injection, $\mathcal{M}_0 = I_0=E_0=0$. Given $Y_1$ and $Y_2$, this null condition identifies the nominal working points of the bridge as a function of $n_1$, $n_2$ and $n_3$, that is, for fixed $Y_1$ and $Y_2$, the null condition with null injection determines a set $\mathsf{Y} = \{Y_3^\textup{n}(n_1,n_2,n_3)\}$ of nominal admittances or, correspondingly, a set $\mathsf{Z} = \{Z_3^\textup{n}(n_1,n_2,n_3)=[Y_3^\textup{n}(n_1,n_2,n_3)]^{-1}\}$ of nominal impedances, for $n_1$, $n_2$ and $n_3$ belonging to $\mathsf{T}$.
\subsection{An example}
\begin{figure}
	\subfigure[$\mathsf{Y}$ locus.]{
%
%
\begin{psfrags}%
\psfragscanon%
%
\psfrag{s03}[t][t]{\color[rgb]{0,0,0}\setlength{\tabcolsep}{0pt}\begin{tabular}{c}$\textup{Re}\,Y_3$\end{tabular}}%
\psfrag{s04}[b][b]{\color[rgb]{0,0,0}\setlength{\tabcolsep}{0pt}\begin{tabular}{c}$\textup{Im}\,Y_3$\end{tabular}}%
%
\psfrag{x01}[t][t]{-1}%
\psfrag{x02}[t][t]{-0.5}%
\psfrag{x03}[t][t]{0}%
\psfrag{x04}[t][t]{0.5}%
\psfrag{x05}[t][t]{\shortstack{1\\$\times 10^{-3}\ $}}%
%
\psfrag{v01}[r][r]{-1}%
\psfrag{v02}[r][r]{-0.5}%
\psfrag{v03}[r][r]{0}%
\psfrag{v04}[r][r]{0.5}%
\psfrag{v05}[r][r]{1}%
\psfrag{ypower1}[Bl][Bl]{$\times 10^{-3}$}%
%
\resizebox{9cm}{!}{\includegraphics{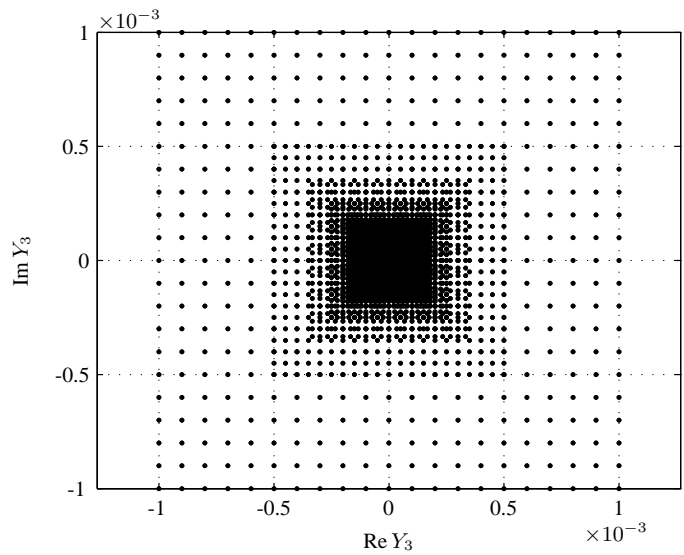}}%
\end{psfrags}%
%
 \label{fig:Yrange}} \\
	\subfigure[$\mathsf{Z}$ locus.]{
%
%
\begin{psfrags}%
\psfragscanon%
%
\psfrag{s03}[t][t]{\color[rgb]{0,0,0}\setlength{\tabcolsep}{0pt}\begin{tabular}{c}$\textup{Re} Z_3$\end{tabular}}%
\psfrag{s04}[b][b]{\color[rgb]{0,0,0}\setlength{\tabcolsep}{0pt}\begin{tabular}{c}$\textup{Im} Z_3$\end{tabular}}%
%
\psfrag{x01}[t][t]{-2}%
\psfrag{x02}[t][t]{-1}%
\psfrag{x03}[t][t]{0}%
\psfrag{x04}[t][t]{1}%
\psfrag{x05}[t][t]{\shortstack{2\\$\times 10^{4}\ $}}%
%
\psfrag{v01}[r][r]{-1.5}%
\psfrag{v02}[r][r]{-1}%
\psfrag{v03}[r][r]{-0.5}%
\psfrag{v04}[r][r]{0}%
\psfrag{v05}[r][r]{0.5}%
\psfrag{v06}[r][r]{1}%
\psfrag{v07}[r][r]{1.5}%
\psfrag{ypower1}[Bl][Bl]{$\times 10^{4}$}%
%
\resizebox{9cm}{!}{\includegraphics{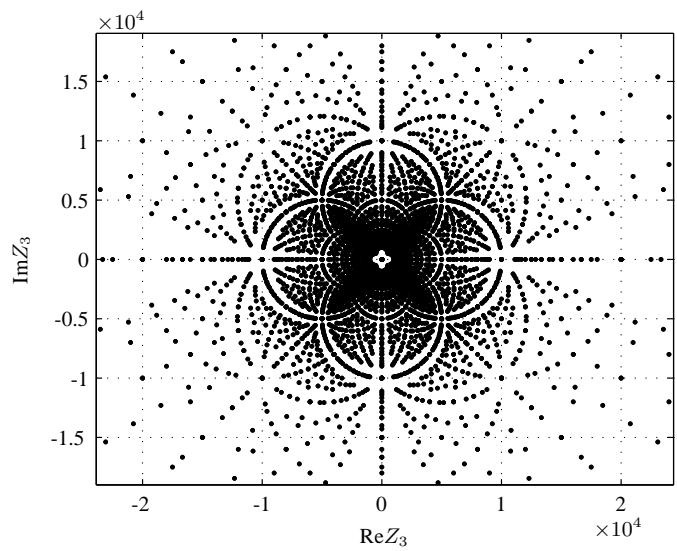}}%
\end{psfrags}%
%
 \label{fig:Zrange}}%
	\caption{An example of the nominal working point set loci of a CC bridge over the complex plane. \subref{fig:Yrange} admittance  $\mathsf{Y}$ locus.  \subref{fig:Zrange} impedance  $\mathsf{Z}$ locus. For clarity, the figures have been zoomed with a different scale factor. \label{fig:range}}%
\end{figure}

\begin{figure}
	\centering
	\includegraphics[width=\linewidth]{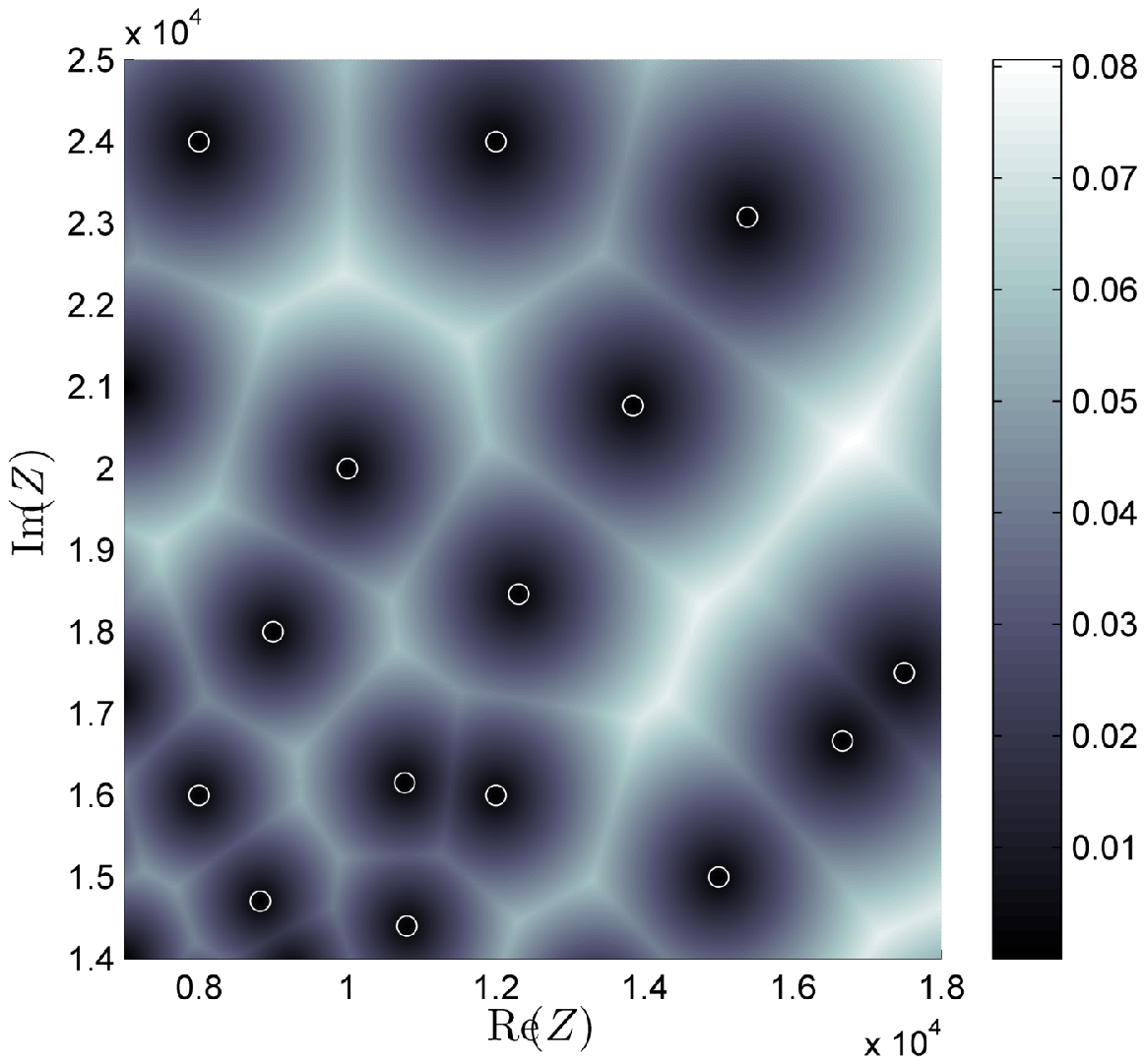}
 	\caption{A zoom view of Fig.~\ref{fig:Zrange}. The $\mathsf{Z}$ locus is represented as white circles; the injection magnitude $|m_0|$ is plotted in grayscale. The Voronoi tessellation of the complex plane can be easily recognized. \label{fig:voronoi}}%
\end{figure}

As an example, let us consider the following set-up:
\begin{itemize}
\item Current comparator having a set of available taps $\mathsf{T}=\left\{ -100, -90, \ldots, 90, 100 \right\}$. The total number of available $(n_1, n_2, n_3)$ tap combinations is $21^3 = 9261$; however, the number of distinct available bridge working points is \num{3216}.
\item $Z_1$ is a pure resistor, $R_1 = \SI{10}{\kilo\ohm}$;
\item $Z_2$ a pure capacitor, $C_2=\SI{10}{\nano\farad}$;
\item The comparison is performed at the frequency $f=\SI{1592}{\hertz}$.
\end{itemize}

Fig.~\ref{fig:range} shows the loci, in the complex plane, of the sets $\mathsf{Y}$ and $\mathsf{Z}$ corresponding to the above given set-up. The particular shape of the $\mathsf{Z}$ locus (Fig.~\ref{fig:Zrange}) can be appreciated by considering that the inversion $Z=1/Y$ is a special case of M\"obius transformation, which maps generalized circles (i.e., including straight lines) to generalized circles.

For given $Y_0,\ldots,Y_3$, the triplet $(n_1,n_2,n_3)$ should be chosen to maximize the detection sensitivity and to minimize the injection $\mathcal{M}_0$ with respect to the other magnetomotive forces involved. A possible criterion is to minimize the absolute value of the quantity $m_0 = \mathcal{M}_0/\lVert M \rVert$, where $\lVert \mathcal{M} \rVert$ is the (quadratic) norm of the magnetization vector $[\mathcal{M}_k]$,
\begin{equation}
	\label{eq:distance}
	\displaystyle m_0  = \frac{\mathcal{M}_0}{\left( \sum_{k=1}^3 |\mathcal{M}_k|^2 \right)^\frac{1}{2}}.  
\end{equation}

Given $Z_1$ and $Z_2$, Eq.~\eqref{eq:distance} can be used to define a metric on the complex plane. This metric induces a corresponding \emph{Voronoi tessellation}~\cite{Dirichlet1850} with convex \emph{Voronoi cells}. A small portion of such tessellation is depicted in Fig.~\ref{fig:voronoi}, which, for each complex $Z_3$, gives also the corresponding magnitude of $m_0$, expressed as a gray scale.  
%
%
\section{A test bridge implementation, digitally-assisted}
\label{sec:implementation}
\begin{figure*}
	\centering
	\def\svgwidth{0.65\linewidth}	
	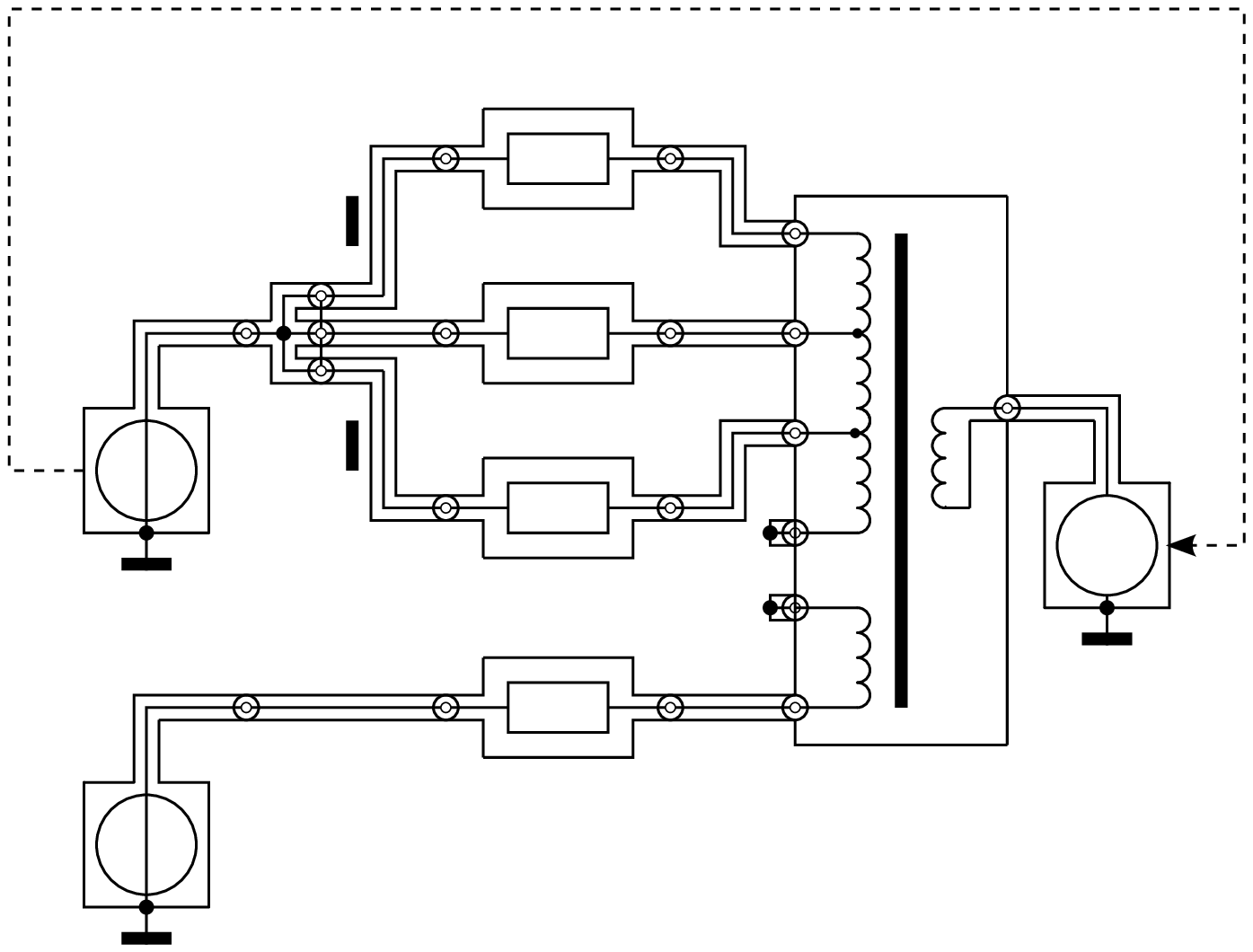
 	\caption{Coaxial schematic diagram of the three-arm current comparator bridge. All impedances are defined as two terminal-pair standards.  \label{fig:coaxscheme}}%
\end{figure*}
\begin{figure}[tb]
	\includegraphics[width=\linewidth]{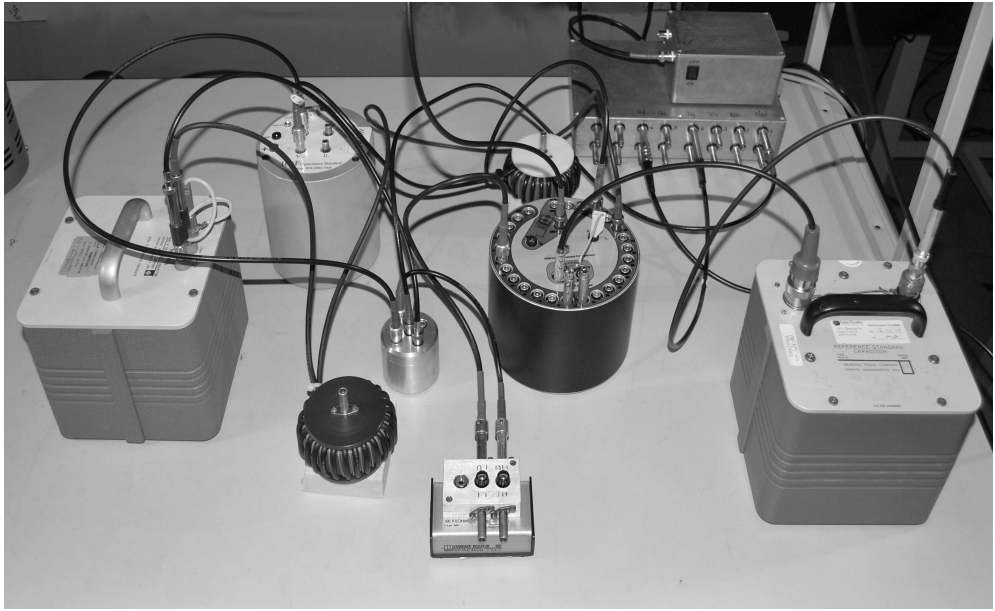}
 	\caption{A photograph of the three-arm current comparator bridge. The CC is the cylindrical object near the figure center.   \label{fig:photo}}%
\end{figure}

\subsection{Impedance definition}
The same schematic diagram of Fig.~\ref{fig:3armCC} can be implemented for different impedance definitions. The test implementation here presented follows the two terminal-pair (2P) definition \cite[Ch. 2]{CallegaroBook}, which achieves high accuracy for mid- and high-value impedances ($|Z| \geq \SI{1}{\kilo\ohm}$). At the expenses of a more complex circuitry
, a four terminal-pair implementation is also possible. 
\subsection{The bridge}
The implementation of the three-arm current comparator bridge here presented is derived from that of a two-arm digitally-assisted current comparator bridge \cite{Trinchera2013}, which showed good results in the comparison of like impedances.  

Digitally-assisted bridges~\cite{Cabiati1985, Ramm1985b, Muciek1997, Corney2003, Trinchera2009, Callegaro2010} are based on the generation of the sine wave signals to be employed in the bridge with digital synthesized sources. This approach permits to simplify the bridge mesh and achieve automated operation. Since the accuracy of a digitally-assisted bridge is granted, like in traditional bridges, by the electromagnetic ratio devices involved, the measurement accuracy is not sacrificed: measurement uncertainties in the \num{E-8} level were demonstrated \cite{Callegaro2010}. 

The coaxial circuit diagram of the bridge is shown in Fig.~\ref{fig:coaxscheme}, and a photograph in Fig.~\ref{fig:photo}.

The electromagnetic current comparator and the polyphase synthesized generator employed in the implementation are described in detail in~\cite{Trinchera2009,Callegaro2010}. The detector employed is a commercial Stanford Research mod.~830 lock-in amplifier. The bridge equilibrium is achieved by adjusting the voltage $E_0$ with a simple automated control strategy~\cite{Callegaro2005}.  
\section{Measurements}
\begin{table*}[tb]

\centering
\caption{Standards employed during the test measurements. \label{tab:standards}}
\setlength{\tabcolsep}{1mm}
\resizebox{\linewidth}{!}{
\begin{tabular}{llll}
\toprule
 & Nominal value & Type & $Y^\textup{ref}$ traceability\\
\midrule
$Y_1$ & \SI{100}{\kilo\ohm} & Electro Scientific Industries mod. SR1, shielded & Italian national ac resistance standard.\\ 
$Y_2$ & \SI{10}{\nano\farad} & Custom realization, thermostated \cite{Callegaro2005b} & Italian national capacitance standard. \\ 
$Y_3^\textup{(a)}$ & \SI{1}{\henry}: quality factor $Q\approx 10$ at \SI{1}{\kilo\hertz} & General Radio mod. 1482-P & Three-voltage method~\cite{Callegaro2001}\\ 
$Y_3^\textup{(b)}$ & $RC$ box ($\approx \SI{100}{\kohm}$, $\SI{5.2}{\nano\farad}$) & Custom realization from Vishay components in a shielded box.& Andeen-Haegarling mod. 2700A\\ 
$Y_0$ & \SI{100}{\pico\farad} & General Radio mod. 1404-A & Italian national capacitance standard.\\
\midrule
\end{tabular}
}
\end{table*}

Preliminary tests of the bridge described in Sec.~\ref{sec:implementation} were performed by selecting nearly-pure impedances, a resistor and a capacitor, for $Y_1$ and $Y_2$, and impure standards for $Y_3$. All impedance magnitudes are in the \SIrange{10}{100}{\kilo\ohm} range at \si{\kilo\hertz} frequency, which allow accurate measurements with two terminal-pair definition. All impedances were separately calibrated with other measurement systems: Tab.~\ref{tab:standards} gives some details about the standards and the calibration traceability route.

Two impure $Y_3$ standards were measured at two frequencies (\SI{1}{\kilo\hertz} and \SI{2}{\kilo\hertz}): $Y_3^\textup{(a)}$ is an air-core inductor $L$, $Y_3^\textup{(b)}$, a parallel $RC$ network. Tab.~\ref{tab:tapset} shows the choice of the CC tap setting $n_k$ for each measurement.

Tab.~\ref{tab:results} reports, for each $Y_3$, the relative deviation ${\Delta y = (Y_3^\textup{bridge}-Y_3^\textup{ref})/Y_3^\textup{ref}}$ between the bridge reading $Y_3^\textup{bridge}$ and the reference value $Y_3^\textup{ref}$ provided by the corresponding calibration. 

The expression of uncertainty of the bridge measurements, which has to be evaluated in the framework of GUM Supplement 2 \cite{GUMS2}, has not been carried out  yet and will be the subject of a separate work. However, the value of the deviation $\Delta y$ can be compared with the calibration uncertainty of $Y_3^\textup{ref}$; in particular, Tab.~\ref{tab:results} reports the calibration uncertainties\footnote{The large values of $u(\varRe Y_3^\textup{ref})$ for the $L$ standard are caused by the large temperature coefficient of the inductor equivalent resistance, of about \SI{3.9E-3}{\per\kelvin}.} $u(\varRe Y_3^\textup{ref})$ and $u(\varIm Y_3^\textup{ref})$ of the real and imaginary parts of $Y_3^\textup{ref}$, relative to its magnitude $|Y_3^\textup{ref}|$. 

The reader can appreciate that the real and imaginary parts of $\Delta y$ are compatible with the corresponding calibration uncertainties for all the measurement performed.
\begin{table}[t]
\centering
\caption{CC taps $n_k$.   \label{tab:tapset}}
\setlength{\tabcolsep}{1mm}
\begin{tabular}{lSSSS}
\toprule
$Y$ & \multicolumn{4}{c}{$n_k$} \\
\midrule
	& \multicolumn{2}{c}{$L$ std} & \multicolumn{2}{c}{$RC$ std} \vspace{1mm}\\
	& $\SI{1}{\kilo\hertz}$ & $\SI{2}{\kilo\hertz}$  & $\SI{1}{\kilo\hertz}$ & $\SI{2}{\kilo\hertz}$ \\ 
\midrule
$Y_1$ &  -60 & -30 & -100 & -80 \\
$Y_2$ &  100 & 50 & -50 & -40 \\ 
$Y_3^{(\textup{a,b})}$ &  40 & 80 & 100 & 80\\ 
$Y_0$ &  20 & 20 & 20 & 20\\
\bottomrule
\end{tabular}
\end{table}
\begin{table}[tb]
\centering
\caption{Measurement results. \label{tab:results}}
\begin{tabular}{ccccc}
\toprule
Std & $f/\si{\kilo\hertz}$ & $\Delta y$ & $u(\varRe Y_3^\textup{ref})/|Y_3^\textup{ref}|$ & $u(\varIm Y_3^\textup{ref})/|Y_3^\textup{ref}|$ \\[1ex]
& & $\times \num{E6}$ & $\times \num{E6}$ & $\times \num{E6}$ \\
\midrule
\multirow{2}{*}{$L$} & $1$ & \num{92-3i} & \num{57} & \num{50} \\
					  & $2$ & \num{19-10i} & \num{28} & \num{50} \\
\multirow{2}{*}{$RC$} & $1$ & \num{-8.1-j6.7} & \num{47} & \num{46} \\
					  & $2$ & \num{-6.6-j3.1} & \num{40} & \num{37} \\
\bottomrule
\end{tabular}
\end{table}
\section{Conclusions}
This paper describes the operating principle of a three-arm current comparator bridge suitable for measuring impedances having arbitrary phase angles. An implementation of the principle has been developed in the form of a digitally-assisted two-terminal pair coaxial bridge; test measurements on an air-core inductor and of an $RC$ network yielded results compatible with previous knowledge on the values of standards involved.

The development of this bridge is part of the European Metrology Research Programme (EMRP) Project SIB53 AIM~QuTE, \emph{Automated impedance metrology extending the quantum toolbox for electricity}. Deliverables of the project include the development of dedicated impure standards and an interlaboratory comparison which will allow the validation of the bridge performance.
 
%
%

\section*{Acknowledgment}
The authors are indebted with Bruno Trinchera, INRIM, for the design of the current comparator employed in the bridge.

The activity has been partially financed by Progetto Premiale MIUR\footnote{\emph{Ministero dell'Istruzione, dell'Universit\`a e della Ricerca.}}-INRIM P4-2011 \emph{Nanotecnologie per la metrologia elettromagnetica} and P6-2012 \emph{Implementation of the new International System (SI) of Units}. The work has been realized within the EMRP Project SIB53 AIM~QuTE, \emph{Automated impedance metrology extending the quantum toolbox for electricity}. The EMRP is jointly funded by the EMRP participating countries within EURAMET and the European Union. 

\bibliographystyle{IEEEtran}
\bibliography{DigitalImpedanceBridges}

\end{document}